\documentclass[aps,prb,twocolumn,superscriptaddress]{revtex4}
\usepackage{graphics}
\usepackage{longtable}
\usepackage[epsfig]{graphicx}
\usepackage{epsfig}
\usepackage{amsmath}

\begin{document}
%

\title{First-principles molecular dynamics simulations of proton diffusion in cubic BaZrO$_3$ perovskite under strain conditions}

\author{Marco \surname{Fronzi}}
\email[Corresponding author\\ E-mail: ]{marco.fronzi@gmail.com}
\affiliation{Department of Mechanical Science and Bioengineering, Graduate School of Science, Osaka University, Osaka, Japan}

\author{Yoshitaka \surname{Tateyama}}
\affiliation{International Center for Materials Nanoarchitectonics (MANA),
National Institute for Materials Science (NIMS), Tsukuba, Japan}

\author{Nicola \surname{Marzari}}
\affiliation{\`{E}cole Polytechnique F\'{e}d\'{e}rale de Lausanne (EPFL), Lausanne, Switzerland}

\author{Enrico \surname{Traversa}}
\affiliation{King Abdullah University of Science and Technology, Physical Sciences and Engineering Division, Kingdom of Saudi Arabia}

\date{\today}

\begin{abstract}

First-principles molecular dynamics simulations have been employed to analyze the proton diffusion in cubic BaZrO$_3$ perovskite at 1300K,
and a non-linear effect of an applied isometric strain of 2$\%$ on the lattice parameter has been observed. The structural and electronic properties of BaZrO$_3$ are analyzed, based on Density Functional Theory calculations, and after an analysis of the electronic structure, we provide a possible explanation for an enhanced ionic conductivity, that can be caused by the formation of a preferential path for proton diffusion under compressive strain conditions.

\end{abstract}





\maketitle


\section{Introduction}

The development of highly ionic conductive materials is particularly important for the fabrication of high-performance solid oxide fuel cells (SOFCs) operating at intermediate/low temperatures (550-900 K). State-of-the-art SOFC technology requires the cell to operate at temperatures between 1000 and 1300 K, which makes both fabrication and operation costly because expensive materials need to be used for sealing and for inter-connectors.\cite{steele2001,aguadero2012} Therefore, to encourage widespread adoption of SOFCs, the development of highly conductive materials that can operate at intermediate/low temperatures is imperative.

\begin{figure}[tbp]
   \begin{center}
   \scalebox{0.23}
   {
   \includegraphics{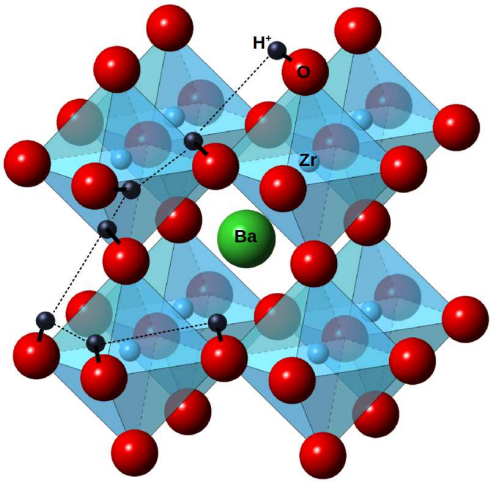} 
   }
   \end{center}
   \caption{ (Color online) Schematic representation of proton diffusion. The H atom bonded to the O atom migrates to the next oxygen atom, breaking a O-H bond and forming a new one. The picture shows the octahedron polyhedrons formed by oxygen atoms, and a possible path for proton diffusion through intra and infra-octahedral jumps.   }
   \label{fig:proton_path}
\end{figure}

 BaZrO$_3$, similar to several other compounds that crystallize in a perovskite structure, has been known to show high proton conductivity and is therefore a good candidate for an electrolyte material capable of operation at the desired operating temperature. \cite{fabbri2009} Ideally, in a perfect perovskite structure, a proton forms an O{-}H bond with one of the four equivalent oxygen atoms, where the O-H group is characterized by vibrational and rotational movement. The proton conduction mechanism has been explained in terms of proton jumps from one oxygen site to another by H transfer and O-H reorientation. \cite{zhang2008, sundell2007} 
Because of the lattice geometry, the possible jumps are classified as intra-octahedral or infra-octahedral. At its stable site, the hydrogen atom strongly interacts with the surrounding atoms; this interaction deforms the lattice by shortening the distance between the proton and the neighboring oxygen. Conduction occurs through a series of transitions between sites coordinated to different oxygens and sites coordinated to the same oxygen (schematically illustrated in Fig. \ref{fig:proton_path}). \cite{cherry1995, gomez2005} 
 It is also well known that the ionic conductivity is typically enhanced by replacing one or more element of the crystal lattice with a lower-valence element.\cite{aguadero2012} The charge redistribution that occurs in order to balance the difference in oxidation state between the doping atom and the substituted atom produces oxygen vacancies. Furthermore, an additional and non-negligible effect of doping is a local distortion and the breaking of the lattice symmetry. This occurs because the doping atom contributes to the creation of new oxygen transport pathways; the local distortion stretches the bond, affecting the activation barriers and the diffusivity of the species.
First-principles molecular dynamics simulations have been widely used to calculate and predict the ion conduction properties of several bulk crystal structures, confirmed the migration mechanism described above.\cite{marrocchelli2009, norberg2009, gomez2007} 

In addition to describing a new class of compounds, the latest studies highlight a great increase in conductivity, that can be achieved by coupling metal-oxide materials having different lattice parameters. This would crate a so-called semi-coherent interface at which the two compounds are subject to strain.\cite{Garcia-Barriocanal2008, korte2008} A semi-coherent interface preserves the crystal structure of the original compound while creating a deformation of the cell, which, however, preserves its original volume. In the cubic perovskite structure, the deformation would result in an interface strain (epitaxial strain), that would produce a tetragonal structure with the same volume as the original cubic one.
Within this framework, it is essential to study the effect of strain on ion diffusion in order to develop a new class of electrolyte materials with good chemical stability and high ion conductivity in the intermediate/low temperature range.

In this work, we analyzed the effect of an external strain on proton diffusion in the cubic Pm-3m perovskite crystal structure of BaZrO$_3$, in order to estimate the conditions under which proton conduction might be enhanced. Here, we considered only isotropic strains in order to avoid the existence of a preferential diffusion direction.
We use a first-principles molecular dynamics approach within the Car-Parrinello approximation, with a temperature condition of 1300 K (fuel cell working conditions). The choice of the temperature has been made to facilitate the diffusion process if compared to the intermediate/low temperature. However, since no physical properties of the lattice would change, the physics behind the results would not change for a choice of temperature between 550 K and 1300 K. 
Also, we are aware that in practical applications, barium zirconate is used in its doped form to facilitate the formation of the oxygen vacancies necessary for proton adsorption and diffusion; however, in this work we analyzed only the pure BaZrO$_3$ crystal because we focused on analyzing the effect of strain on proton diffusion through oxygen atoms.

\section{Calculation Methods}

\subsection{conditions for electronic states calculations}

We used plane-wave basis density functional theory (DFT) as implemented in the Car-Parrinello code of the Quantum-ESPRESSO distribution.\cite{giannozzi2009} We used a Perdew-Burke-Ernzerhof functional for the exchange-correlation term.\cite{perdew1996} Ultrasoft pseudo-potentials were used to simulate the effect of the core electrons. Twelve valence electrons in the 6s$^2$5p$^6$5s$^2$ and 5s$^2$4d$^2$4p$^6$4s$^2$ configurations were considered for barium and zirconium atoms, respectively, and four electrons in 2p$^4$2s$^2$ were considered for oxygen atoms. The cut-off energies for the wave function and charge density were 27.0 Ry and 240 Ry, respectively.

\subsection{Conditions for dynamics calculations}

\begin{figure}[tbp]
   \begin{center}
   \scalebox{0.27}
   {
   \includegraphics{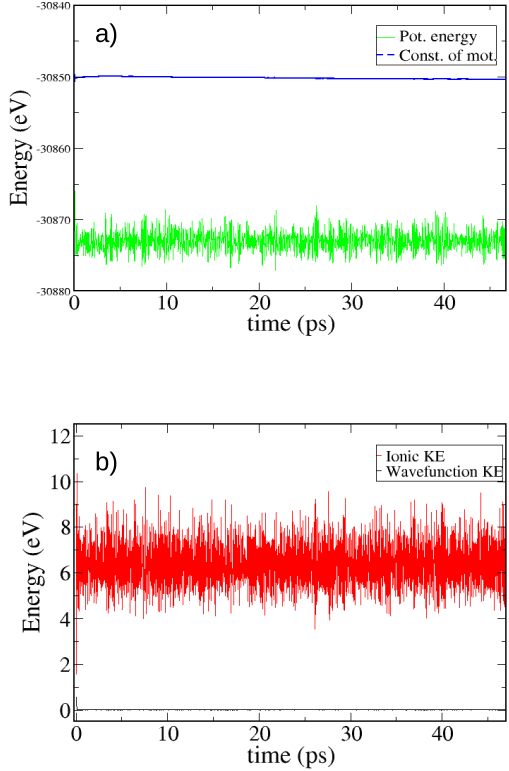} 
   }
   \end{center}
   \caption{ (Color online) (a) Potential energy and constant of motion (blue and green lines, respectively), and (b) Ionic and wave-function kinetic energy (red and black lines, respectively),
   calculated in a 40 atom supercell, with a choice of the fictitious electron mass of 150 a.u. and of the time step dt = 3.87 fs. The temperature of the simulation was maintained at 1300 K by a Nos\'{e}-Hoover thermostat}
   \label{fig:ion-wfc-ene-sys}
\end{figure}

We performed Car-Parrinello molecular dynamics simulations using a fictitious mass corresponding to a 150 a.u. wave function. The simulations were performed in a supercell of 40 atoms (symmetry group Pm-3m) and in the canonical ensemble. To reproduce the working conditions of a fuel cell, the temperature of the simulations was maintained at 1300 K by a Nos\'{e}-Hoover thermostat. 
The runs, each lasting for more than 40 ps, were performed using a fictitious electron mass m =150 a.u. and a time step dt = 3.87 fs. These choices allow for excellent conservation of the constant of motion (Fig. \ref{fig:ion-wfc-ene-sys}a shows negligible dissipation during a 46 ps simulation) and negligible drift in the fictitious kinetic energy of the electrons for simulations of that duration. In addition, the ratio between the kinetic energies of ions and electrons was R $<$ 1/20 for the entire simulation time, as shown in Fig. \ref{fig:ion-wfc-ene-sys}b. The separation between the ionic and wave-function kinetic energies in Fig. \ref{fig:ion-wfc-ene-sys}b indicates a good approximation of the fictitious electron mass. We discarded from each trajectory the initial 4 ps, during which the ions reached their target kinetic energy.


\begin{figure}[tbp]
   \begin{center}
   \scalebox{0.23}
   {
   }
   \end{center}
   \label{fig:system-energy}
\end{figure} 

\subsection{Hydrogen charge treatment}

Bj\"{o}rketun {\em et al.} studied the charge state of the hydrogen atom.\cite{bjorketun2007} Using DFT calculations, they estimated the interstitial hydrogen atom formation energy as a function of the Fermi energy. Their results showed that the hydrogen atom is stable in the charge state +1 (the proton) for every value of the Fermi energy within the band gap range. \cite{bjorketun2007} Therefore, in this work we decided to consider only the diffusion of the positive charge state of the hydrogen (the proton).
We obtained a value of 0.18 eV for the proton defect formation energy in a relaxed supercell (8 cells) calculated when the Fermi level is at the valence band maximum. For BaZrO$_3$ cubic perovskite, this value has been calculated for both relaxed and unrelaxed supercells as 0.05 eV (in a 2$\times$2$\times$2 supercell) and 0.21 eV (in a 3$\times$3$\times$3 supercell); the same energy becomes 0.95 eV (in a 2$\times$2$\times$2 supercell) and 1.35 eV (in a 2$\times$2$\times$2 supercell) in an unrelaxed cell. \cite{bjorketun2007}


\section{Results}

\begin{figure}[tbp]
   \begin{center}
   \scalebox{0.28}
   {
   \includegraphics{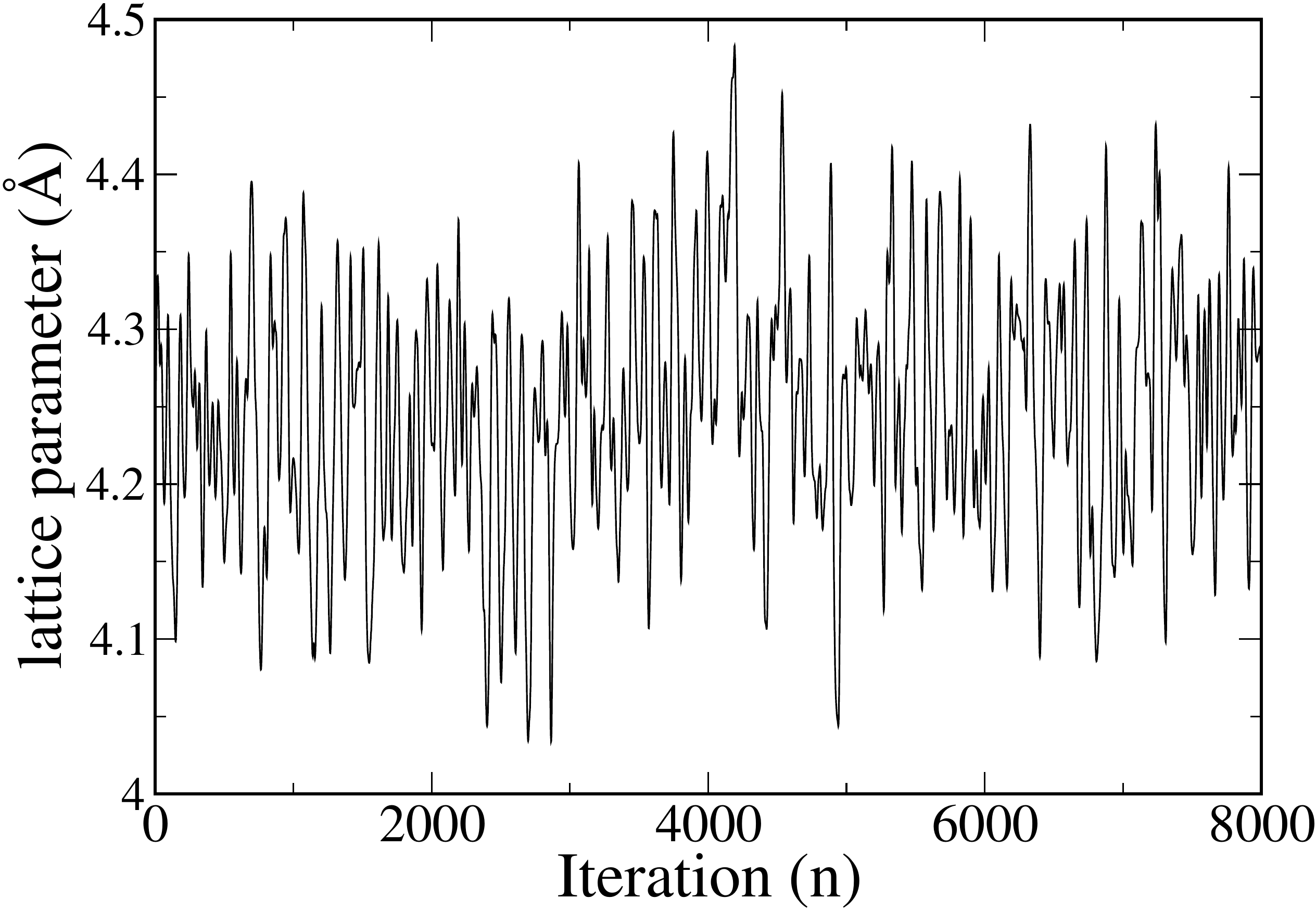} 
   }
   \end{center}
   \caption{ (Color online) Lattice constant of a variable supercell as a function of simulation time step calculated at 1300 K.}
   \label{fig:latt-const-var-cell}
\end{figure} 

We obtained a value of 4.21 {\AA} for the relaxed lattice constant of the stoichiometric BaZrO$_3$ bulk by cell optimization at T = 0 K; this is consistent with other DFT studies (4.20 {\AA}) and with the experimental value (4.19 {\AA}).\cite{shi2005,pagnier2000} 
After a proton is introduced into the bulk structure, in the three cases of tensile, relaxed and compressed lattice, structural relaxation yields an O-H bond distance of 0.98 {\AA}, according to the geometry optimization.
 The positive charge space region in the area surrounding the proton produces a structural distortion compared to the stoichiometric BaZrO$_3$ bulk with a Zr-O-Zr angle of 162.55$^{\circ}$ (180.00$^{\circ}$ in the stoichiometric bulk).

We analyzed the effect of temperature on the lattice parameter expansion by calculating the average change in the cell parameter as a function of time at T = 1300 K, as illustrated in Fig. \ref{fig:latt-const-var-cell}. 
We simulated the BaZrO$_3$ stoichiometric bulk using a variable cell size, starting with a thermally expanded lattice parameter as reported by experimental measurements, in order to estimate the effect of the temperature on the lattice parameter and estimate the relative expansion coefficient.\cite{zhao1991} The calculated average lattice constant in a variable cell simulation at T = 1300 K was 4.236 {\AA}, and the calculated expansion coefficient was Ec = 5.4 $\times$ 10$^{-6}$ K$^{-1}$, whereas the experimental data in the literature report a value of 7.8 $\times$ 10$^{-6}$ K$^{-1}$. \cite{zhao1991}
To simulate an applied external strain (either tensile or compressive), we also applied a further variation of 2$\%$ (0.1 {\AA}) to the calculated and thermally expanded lattice parameters. The simulations were run at a constant lattice parameter for the relaxed bulk and the bulk under compressive or tensile strain.

\begin{figure}[tbp]
   \begin{center}
   \scalebox{0.28}
   {
   \includegraphics{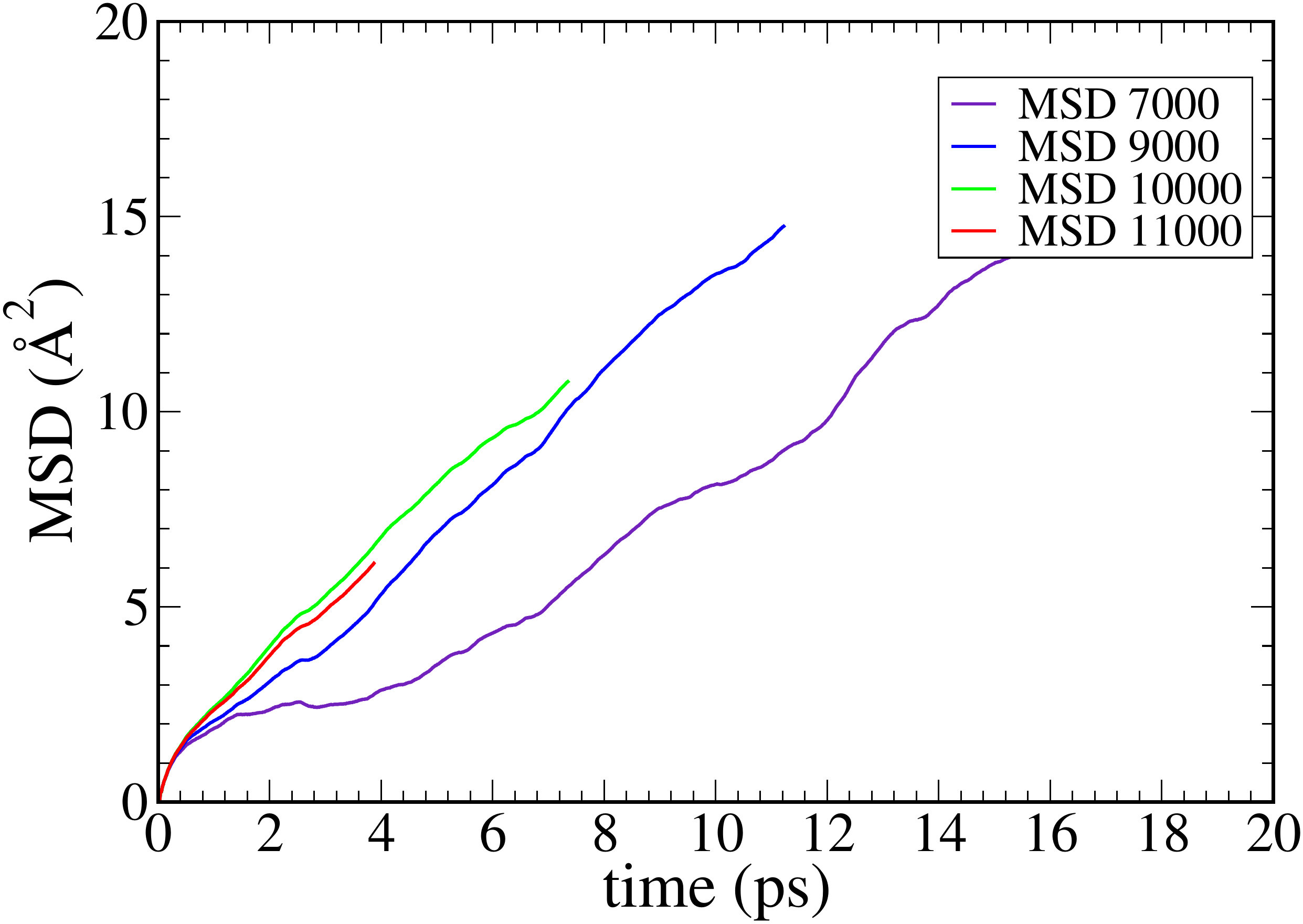} 
   }
   \end{center}
   \caption{ (Color online) Mean square displacement of the proton during self-diffusion calculated in the fully relaxed BaZrO$_3$ bulk at 1300 K averaged over various numbers of time steps.}
   \label{fig:MSD}
\end{figure}

For each condition, we calculated the mean square displacement (MSD) of the proton during a simulation of 46 ps. We considered the average value of the MSD of the proton during self-diffusion over different durations, as shown in Fig. \ref{fig:MSD} for the relaxed bulk. An average over N = 11,000 time steps yielded a good approximation of a linear relationship between the MSD and the time.

Next, we calculated the self-diffusion coefficient (D) from the MSD using the Einstein relation:

\begin{equation}
6D_{self}= \lim_{t\to\infty}\frac{d}{dt} \left\langle | r_{i}(t)-r_{i}(0) | ^{2} \right\rangle ,
\end{equation}

where r is the position of the proton at each time step t. In the relaxed bulk, the calculated diffusion coefficient was 2.3 $\times$ 10$^{-5}$ cm$^2$/s. This value is slightly overestimated compared with other computational work in the literature, which reported a calculated diffusion coefficient of about 1.5 $\times$ 10$^{-5}$ cm$^2$/s for the relaxed bulk at 1000 K. \cite{raiteri2011} However, our value lies between the experimental values reported in the same paper.

\begin{figure}[tbp]
   \begin{center}
   \scalebox{0.28}
   {
   \includegraphics{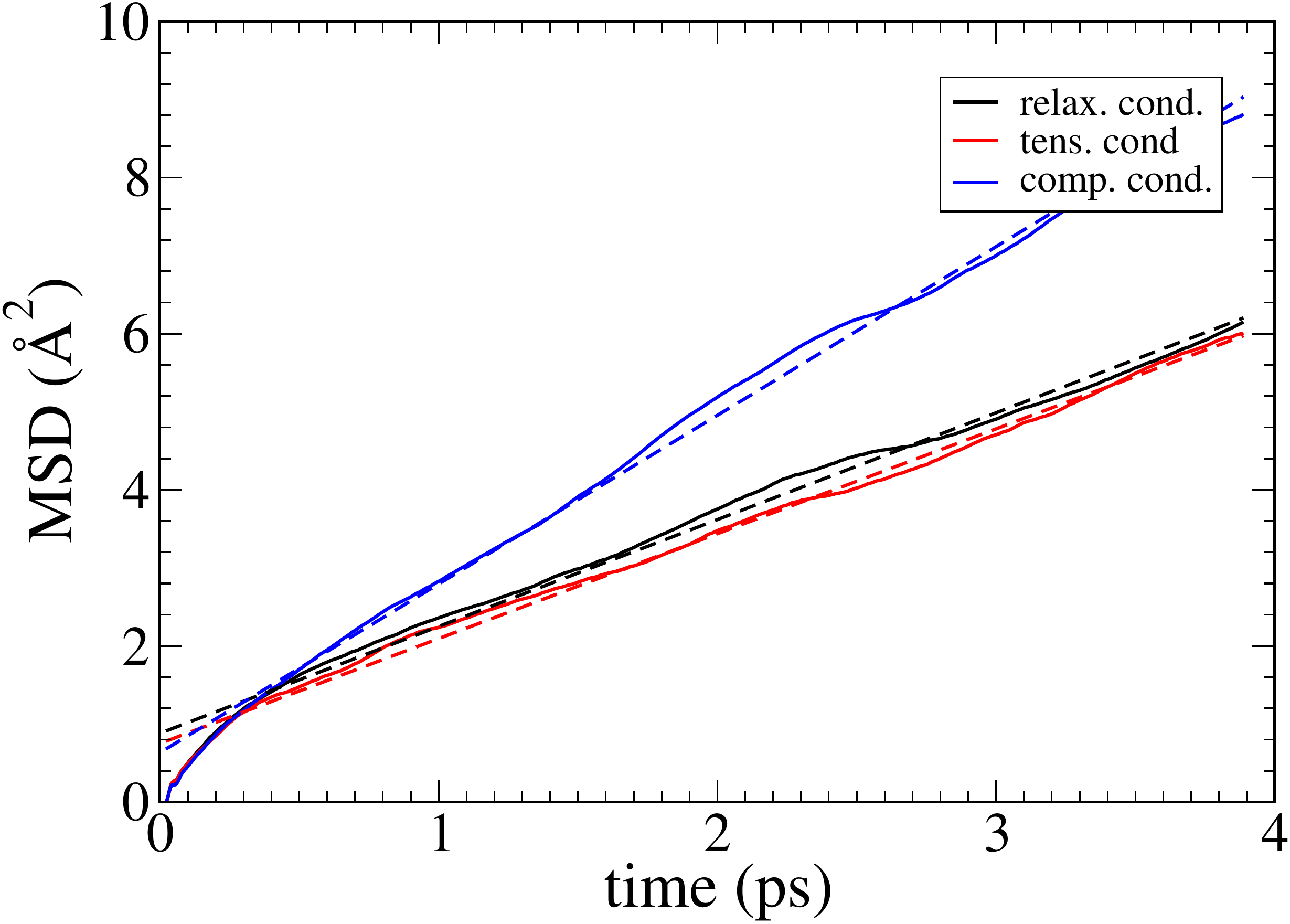} 
   }
   \end{center}
   \caption{ (Color online) Mean square displacement of proton during diffusion calculated in the BaZrO$_3$ bulk under fully relaxed, isometric tensile strain, and compressive strain conditions at 1300 K (black, red, and blue lines, respectively).}
   \label{fig:prot-diff}
\end{figure} 

The D values calculated for the bulk under tensile and compressive strain were 2.2 $\times$ 10$^{-5}$ cm$^2$/s and 3.5 $\times$ 10$^{-5}$ cm$^2$/s, respectively. The proton diffusivity under compressive strain is clearly enhanced compared with the other two conditions. These results show that the change in the lattice parameter under strain (in this case, an equal change in length is made in the relaxed lattice value) does not result in a linear variation of the proton diffusivity. By way of explanation, when a tensile strain is applied to the relaxed bulk, D decreases very slightly, whereas a compression of the same length clearly increases D. 

\begin{figure}[tbp]
   \begin{center}
   \scalebox{0.28}
   {
   \includegraphics{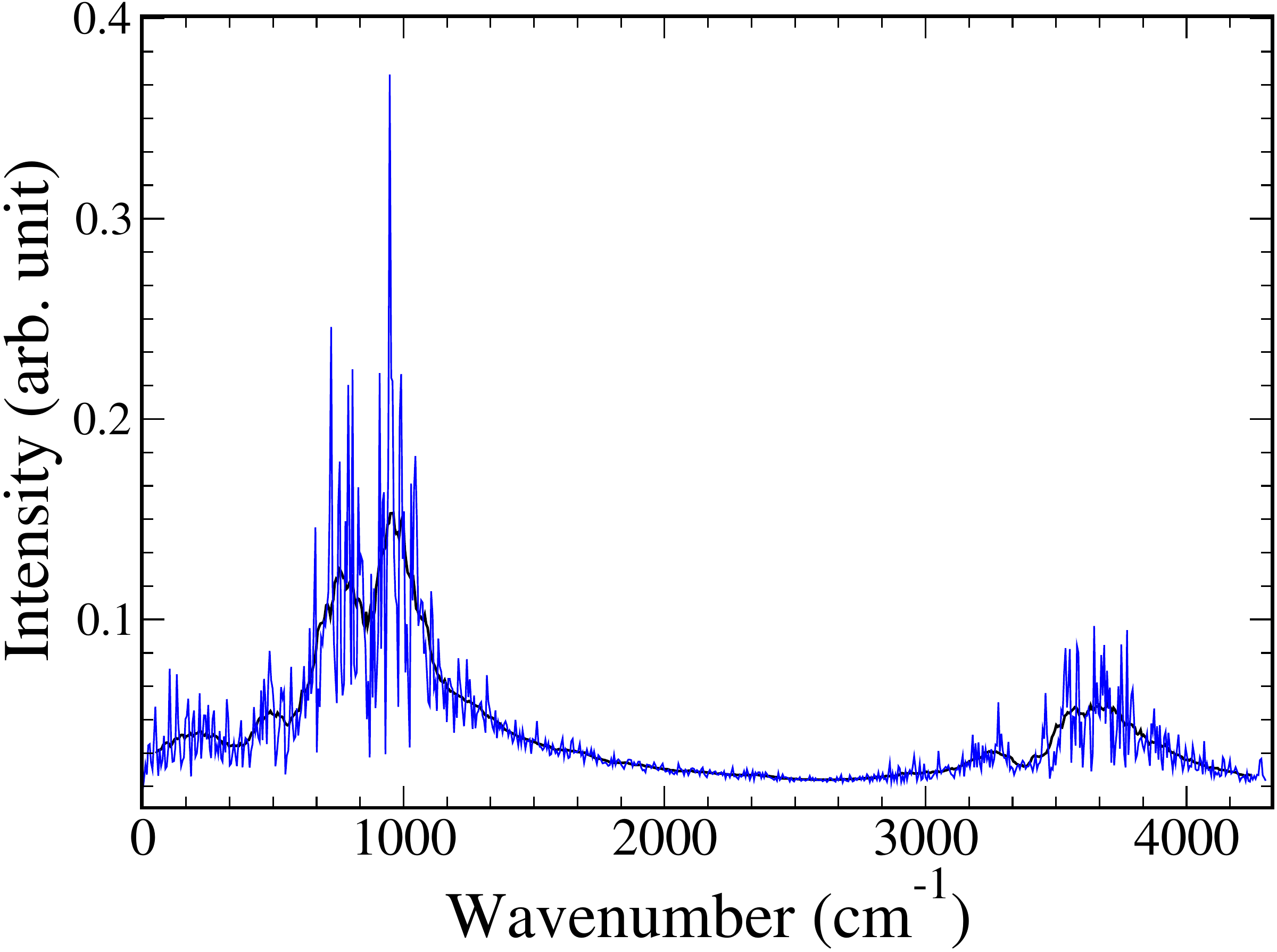} 
   }
   \end{center}
   \caption{ (Color online) Power spectrum of calculated proton diffusion in the fully relaxed BaZrO$_3$ bulk. Spectrum was obtained by Fourier transform of velocity-velocity correlation function. One peak might resemble that of the O-H bond in water at around 3600 cm$^{-1}$.}
   \label{fig:freq_h}
\end{figure} 

To understand the origin of this effect, we analyzed the typical vibration frequencies of the proton, the typical O-H distances, and the electronic structure of the system under different strain conditions. 
We analyzed the typical vibration frequency of the proton by performing a Fourier transform of the velocity-velocity correlation function (Fig. \ref{fig:freq_h}). The typical frequencies have two main peaks at 700-900 cm$^{-1}$ and 3500-3700 cm$^{-1}$. The former may represent the frustrated reorientation of the O-H axis, and the latter may represent the O-H stretching vibration. Both peaks are in good agreement with infrared spectroscopy analysis; however, no substantial difference was observed under the considered strain conditions. \cite{kreuer1999}

\begin{figure}[tbp]
   \begin{center}
   \scalebox{0.28}
   {
   \includegraphics{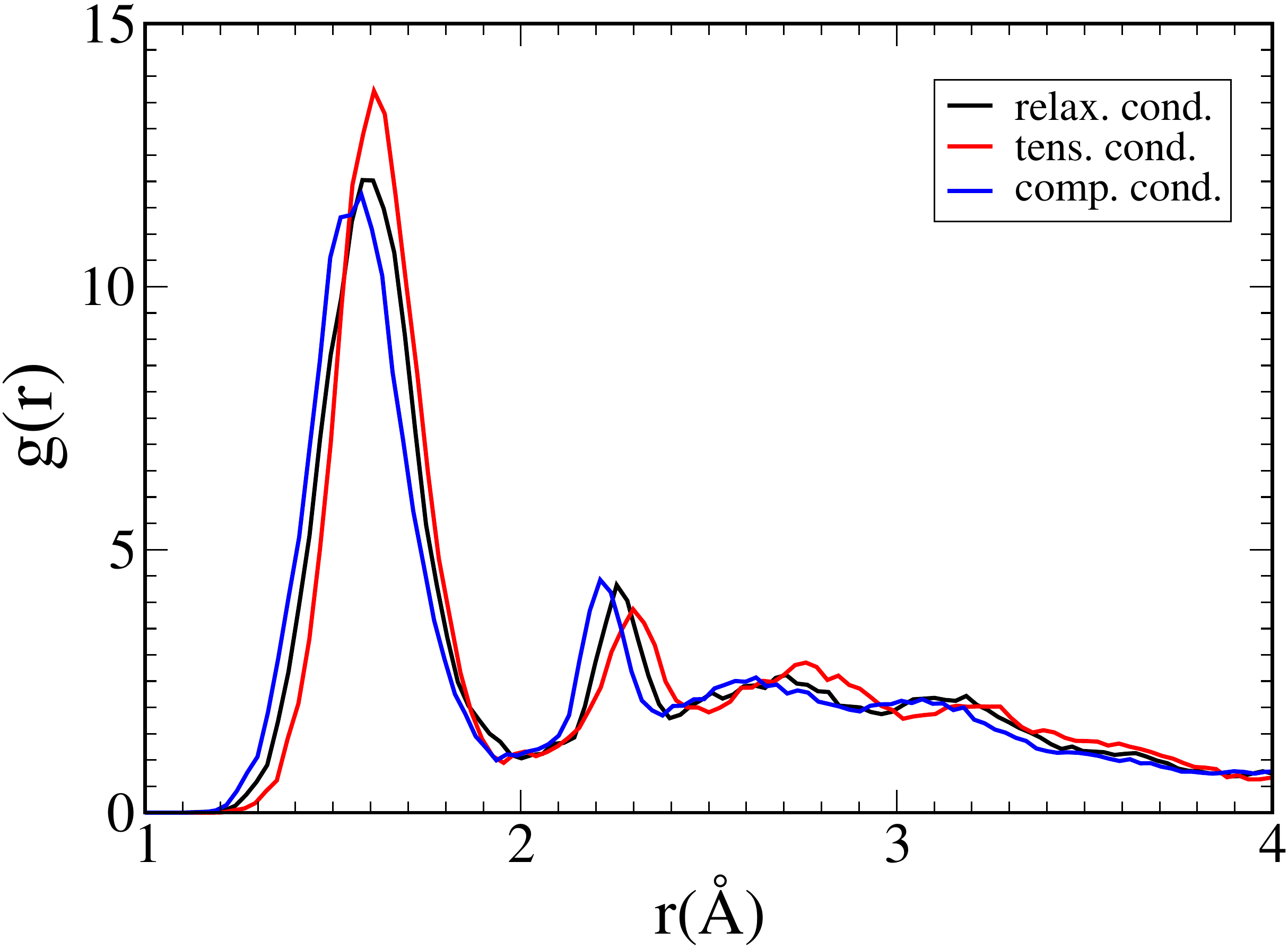} 
   }
   \end{center}
   \caption{ (Color online) Calculated pair correlation function of oxygen-oxygen atoms in the BaZrO$_3$ bulk under relaxed, tensile strain, and compressive strain conditions (red, blue, and black lines, respectively).}
   \label{fig:gr_OO}
\end{figure} 

\begin{figure}[tbp]
   \begin{center}
   \scalebox{0.28}
   {
   \includegraphics{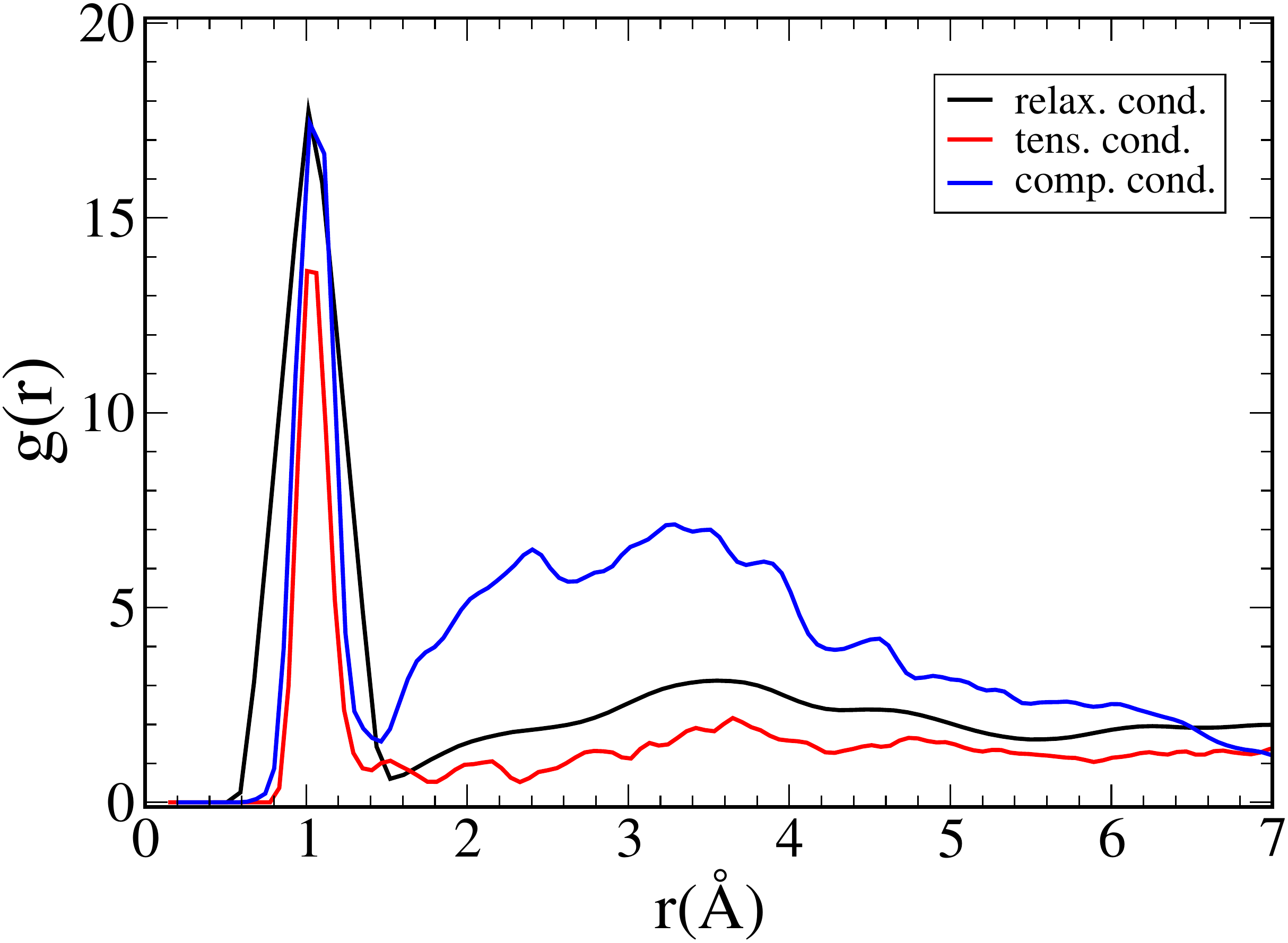} 
   }
   \end{center}
   \caption{ (Color online) Calculated pair correlation function of proton-oxygen atom in the BaZrO$_3$ bulk under relaxed, tensile strain, and compressive strain conditions (red, blue, and black lines, respectively).}
   \label{fig:gr_OH}
\end{figure} 

 We calculated the radial distribution function ($g(r)$) in order to evaluate how the probability of finding an oxygen atom changes with the distance from the proton and with the distance from another oxygen atom.  The three cases show a negligible difference in the oxygen-oxygen radial distribution, with a clear peak at 1.6 {\AA} and a second peak around 2.2-2.3 {\AA} (Fig. \ref{fig:gr_OO}). On the other hand, in the proton-oxygen distribution, a clear peak appears at around 1 {\AA}, indicating oxygen binding; a second peak appears at different locations in the three cases. When the bulk is fully relaxed, the pair correlation function shows a broad peak around 3.12-3.16 {\AA}, although this peak is less evident in the bulk under tensile strain. In the bulk under compressive strain, a pronounced peak appears around 3.12-3.16 {\AA}, with a third peak appearing around 2.4 {\AA} (Fig. \ref{fig:gr_OH}). 
Since the second-third peak is broad and shows a very high intensity, it indicates a higher probability of finding a second oxygen atom close to the proton, suggesting the origin of a further O-H interaction with a second O (O$_B$ in Fig. 1) in addition to the structural O-H bond (O$_A$ in Fig. 1). 

\begin{figure}[tbp]
   \begin{center}
   \scalebox{0.23}
   {
   \includegraphics{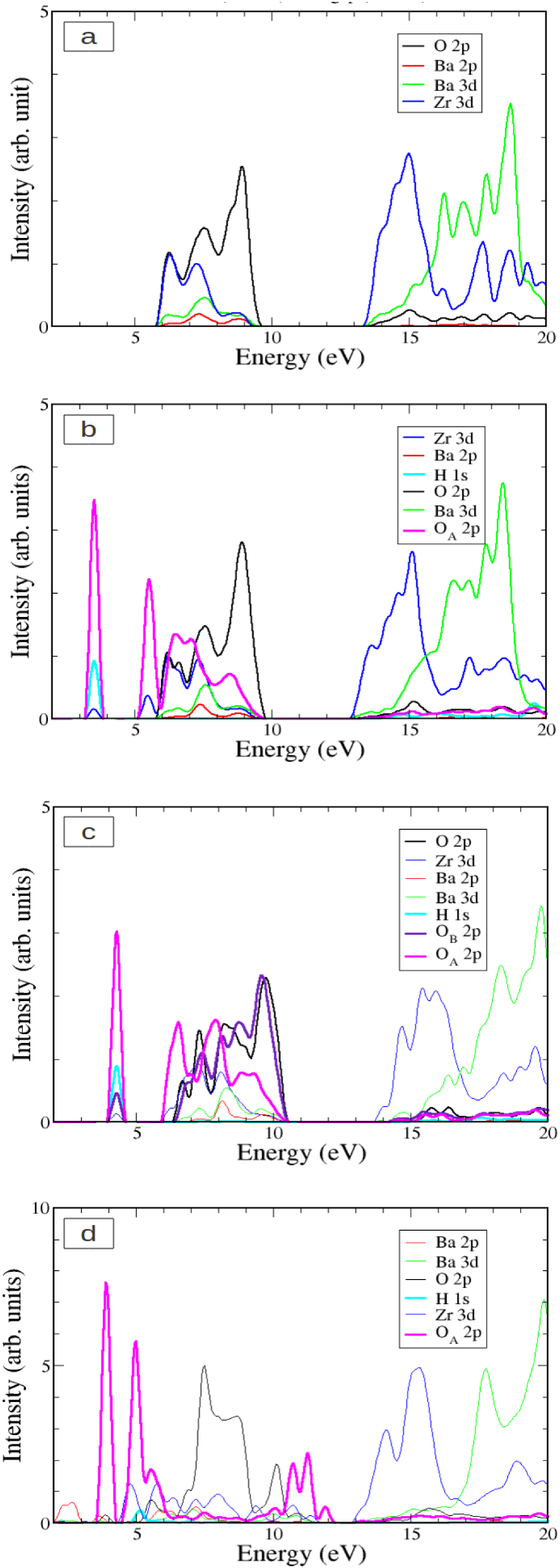} 
   }
   \end{center}
   \caption{ Projected density of states (PDOS) of BaZrO$_3$ (a) under relaxed stoichiometric conditions, (b) after introduction of a proton under relaxed conditions, and after introduction of a proton under (c) compressive and (d) tensile strain.}
   \label{fig:pdos-tot}
\end{figure} 
We found a substantial difference in the electronic structures of the relaxed, compressed, and strained bulk. The introduction of the proton breaks the symmetry of the oxygen atoms Kohn-Sham states, and the projected density of states (PDOS) shows a strong O$_A$-H bond (Fig. \ref{fig:pdos-tot}b) formed by hybridization of the H$_{1s}$ and O$_{2p}$ states and peaking around 3.5 eV. In addition, a second peak representing the oxygen 2p state appears around 5.5 eV. When the bulk is under compressive strain, the PDOS shows further hybridization of the H$_{1s}$ with a second (next-nearest O$_B$ in Fig. \ref{fig:pdos-tot}c) oxygen atom around 4-4.5 eV, suggesting the origin of a new, less pronounced orbital hybridization with another oxygen atom. In the bulk under tensile strain, this type of hybridization did not appear (Fig. \ref{fig:pdos-tot}d).  This suggest the formation, in compressed conditions, of a hydrogenic bond not present in the original structures that may facilitate proton migration by providing a path for proton diffusion.

\begin{figure}[tbp]
   \begin{center}
   \scalebox{0.23}
   {
   \includegraphics{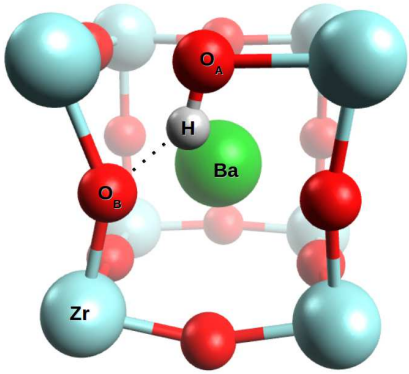} 
   }
   \end{center}
   \caption{ (Color online) Schematic representation of proton diffusion. The H atom bonded to the O$_A$ atom migrates to the next oxygen atom, breaking the O$_A$-H bond and forming a new O$_B$-H bond}
   \label{fig:proton_path_detail}
\end{figure} 

\begin{figure}[tbp]
   \begin{center}
   \scalebox{0.23}
   {
   \includegraphics{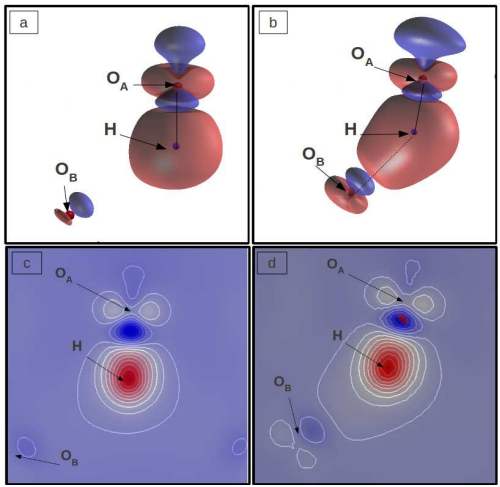} 
   }
   \end{center}
   \caption{ (Color online) Induced charge density due to introduction of a proton in the compressed (b and d) and relaxed (a and c) BaZrO$_3$ bulk. Red areas indicate charge accumulation; blue areas indicate charge depletion.}
   \label{fig:bond-no-bond}
\end{figure} 


The formation a hydrogenic bond is also suggested by the O$_B$-H distances for the next-nearest oxygen atom O$_B$ in the three structures. By using plane DFT calculations, we optimized the three structures and found an O$_B$-H distance of 2.13 {\AA} in the fully relaxed bulk and in the bulk under tensile strain. Thus, the geometry of the structure under tensile strain is qualitatively similar to that of the relaxed bulk. This distance became 1.63 {\AA} in the compressed bulk, which is comparable to the hydrogenic O-H bond distance in liquid water.

Finally, we analyzed the charge redistribution after protonation of the bulk structures. We calculated the charge redistribution by taking the difference between the charge distributions of the stoichiometric and protonated bulk. In the compressed bulk, we found a quasi-symmetric charge distribution around the proton in the direction of two neighboring oxygens, suggesting the presence of a second (weaker) O-H bond, in addition to the structural O-H bond (Figs. \ref{fig:bond-no-bond}b and \ref{fig:bond-no-bond}d). This symmetry did not appear in the other structure (Figs. \ref{fig:bond-no-bond}a and \ref{fig:bond-no-bond}c) and might indicate the formation of a natural pathway in the compressed structure that facilitates proton diffusion (schematically represented in Fig.\ref{fig:proton_path_detail}). 


\section{Conclusion}

We investigated proton diffusion in the BaZrO$_3$ cubic perovskite bulk crystal. We analyzed proton diffusion under fully relaxed conditions as well as under tensile and compressive strain, and the MSD indicates that the strain has a non-linear effect on the proton diffusion constant. There is an evident enhancement under compressive strain, whereas there is no sensible difference between the relaxed bulk crystal and that under tensile strain. The power spectrum obtained by a Fourier transform of the velocity-velocity autocorrelation function showed two main peaks, one of which (3600 cm$^{-1}$) is likely to indicate the O-H stretching mode. However, no obvious differences appeared under either tensile or compressive strain. We also calculated the pair correlation functions of O-O and O-H in the bulk and found a significant change in the latter under compression compared to the other two conditions, suggesting the origin of a further O-H interaction, with a second O in addition to the structural O-H bond. 
The PDOS indicates a clear change in the electronic structure of the compressed protonated bulk compared to the stoichiometric bulk, the protonated relaxed bulk, and the bulk under tensile strain, confirming the formation of a second O-H bond. An analysis of the charge redistribution after the introduction of a proton into the structures also indicates the presence of a second, weak O-H bond in addition to the structural O-H bond, but only in the compressed structure. This indicate the formation of a natural pathway in the compressed structure that facilitates proton diffusion.






\bibliography{bazro3}

\end{document}